\documentclass[aps,prd,preprint,groupedaddress]{revtex4-1}

\usepackage{graphicx} 

\begin{document}

\title{ Thermal width of heavy quarkonia from an AdS/QCD model  }



\author{Nelson R. F. Braga}\email{braga@if.ufrj.br}
\affiliation{Instituto de F\'{\i}sica,
Universidade Federal do Rio de Janeiro, Caixa Postal 68528, RJ
21941-972 -- Brazil}

\author{Luiz F.  Ferreira  }\email{luizfaulhaber@if.ufrj.br}
\affiliation{Instituto de F\'{\i}sica,
Universidade Federal do Rio de Janeiro, Caixa Postal 68528, RJ
21941-972 -- Brazil}


\begin{abstract} 
 We estimate the thermal width of a heavy quark anti-quark pair inside a strongly coupled plasma using a holographic AdS/QCD model. 
 The  imaginary part of the quark potential that produces the thermal width appears in the gravity dual from  quantum fluctuations of the string world sheet in the vicinity of the horizon. The results, obtained using a soft wall background that involves an infrared mass scale,  are consistent with previous analisys where the mass scale was introduced by averaging over 
 quark anti-quark states.

\end{abstract}


\maketitle
  
 \section{Introduction}
 Gauge string duality\cite{Maldacena:1997re,Gubser:1998bc, Witten:1998qj}
 provides an important tool to calculate properties of gauge theores at strong coupling.  
 One quantity of particular interest is the static rectangular Wilson loop, that provides the potential energy
 between an infinitely heavy quark anti-quark pair.  In refs. \cite{Maldacena:1998im,Rey:1998ik} it was proposed that a
 Wilson loop for a gauge theory (at large number of colors and with extended supersymmetry)   is dual to a string worldsheet in  anti-de Sitter space whose boundary is the loop. 
 
 Static Wilson loops can also be calculated for gauge theories at finite temperature.  The energy obtained this way can be taken as the heavy quark potential at finite temperature.  This potential has in general an imaginary part associated with the thermal decay, as discussed for example in \cite{Laine:2006ns, Beraudo:2007ky,Brambilla:2008cx}. 
  The holographic description of Wilson loops in the finite temperature case was developed in \cite{Rey:1998bq,Brandhuber:1998bs}. In this case  the dual geometry is an anti-de sitter black hole.
 
 More recently, the presence of an imaginary part in the quark anti-quark potential was investigated using gauge string duality in, for example,  refs.   \cite{Albacete:2008dz,Noronha:2009da,Giataganas:2013lga,
Finazzo:2013aoa,Fadafan:2013bva,Hayata:2012rw}. In particular, in refs. \cite{Noronha:2009da,Finazzo:2013aoa}  the imaginary part of the quark anti-quark potential is used to calculate the thermal width of a quarkonium state. The state is represented by a static string in a black hole AdS space with end-points fixed on the boundary. The imaginary part of the potential comes from fluctuations of the string near the horizon.
 The thermal width is calculated as the expectation value of the  imaginary part of the potential in a state of the quarkonium.   In this approach of \cite{ Finazzo:2013aoa}  the dual geometry does not contain any dimensionfull parameter (mass scale). The geometry is just a black hole AdS (Poincar\'e)  space, that is dual to a conformal gauge theory. 
 The mass scale of the quarks enters into the calculation of the thermal width through the introduction of a wave function  representing a massive quark subject to a coulomb like potential.  
 
 Here we present an alternative holographic approach to determine the thermal width of a quarkonium state. 
 The motivation is that  AdS/QCD models, like hard wall   \cite{Polchinski:2001tt,BoschiFilho:2002ta,BoschiFilho:2002vd}    and soft wall  \cite{Karch:2006pv} and improved holographic QCD \cite{Gursoy:2007cb,Gursoy:2007er,Gursoy:2009jd}, provide a nice  phenomenological description of quark anti-quark interaction and other hadronic properties like the mass spectra.
 The potential at zero temperature is linearly confining  while  at finite temperature  they exihibit a confinement/deconfinement phase transition both in the hard wall\cite{BoschiFilho:2005mw,BoschiFilho:2006pe} and in the soft wall with positive exponential    factor\cite{Andreev:2006nw}.  We will show here that such a phenomenological descrition of
 quark anti-quark interaction provides also a tool for calculating the thermal width. 
  
 We will consider a quark anti-quark pair in the soft wall model background, that involves an infrared energy scale 
 associated with the mass. The thermal width will be calculated by just averaging over the lengths of all possible string worldsheet configurations that generate imaginary contributions to the potential.    
 For completeness, we mention that  meson widths have been calculated in the holographic D7 brane model framework in \cite{Faulkner:2008qk}.

\section{Holographic description of quark anti-quark potential}

Following the standard gauge/gravity prescription\cite{Maldacena:1998im,Rey:1998ik}  the expectation value of a static
Wilson loop W(C) in a strongly coupled gauge theory that has a gravity dual is represented by the generating functional
$Z_{str}$ of a static string in the bulk of the dual space. The intersection of the string worldsheet with the boundary of the space is the loop C. In the semi-classical gravity approximation we have 
\begin{equation}\label{t2}
 Z_{str}\sim e^{iS_{str}} \,,
\end{equation}
where $S_{str}$ is the classical Nambu-Goto action
\begin{equation}\label{t3}
S_{str}=S_{NG}=\frac{1}{2\pi \alpha'}\int d\sigma d\tau \sqrt{ det(G_{\mu\nu}\partial_{a}X^{\mu}\partial_{b}X^{\nu})},
\end{equation}
where $X^{\mu}(\tau,\sigma)$ are the worldsheet embedding coordinates, $\mu,\nu=0,1,...,4\, $; $a,b=\sigma,\tau$;
$1/2 \pi \alpha'$ is the string tension and $G_{\mu\nu}$ the spacetime metric that we consider to be of Euclidean form. 

A systematic analysis of static strings representing Wilson loops was presented in ref. \cite{Kinar:1998vq}, assuming metrics of  the general form 
\begin{equation}\label{t4}
ds^2=G_{00}(z)dt^{2}+G_{\vec{x}\vec{x}}(z)d\vec{x}^{2}+G_{zz}(z)dz^2 \,,
\end{equation}
where $\vec{x} $ denotes the usual spatial boundary coordinates while $z$ is the radial direction. For our case of interest   the boundary is assumed to be at $z \rightarrow 0$. Choosing the world sheet coordinates $\sigma = x$ and $\tau=t$ and assuming translation invariance along $t$, the string action with endpoints fixed at $x = \pm L/2 $ takes the form
\begin{equation}\label{t5}
S_{NG}=\frac{\mathcal{T}}{2\pi\alpha'}\int^{L/2}_{-L/2}dx \sqrt{M(z(x))(z'(x) )^2 +V(z(x))}
\end{equation}
where 
\begin{eqnarray}
M(z) &\equiv&  G_{00}G_{zz} \label{M}\\
V(z) &\equiv &  G_{00}G_{xx}  \label{V}
\end{eqnarray} 
The string profile $z(x) $ can be determined by considering expression  (\ref{t5}) as representing an ``action integral'' for the evolution in coordinate x.  The corresponding lagrangian density is
\begin{equation}\label{t6}
\mathcal{L}( z, z' )=\frac{1}{2\pi \alpha'} \,\sqrt{M(z) \,z'^2+V(z)\, }
\end{equation}
with  conjugate momentum:
\begin{equation}\label{t7}
p=\frac{\partial\mathcal{L}}{\partial z'}= \frac{1}{2\pi \alpha'} \, \frac{M(z)z'}{\sqrt{M(z)z'^2+V(z)}} \,,
\end{equation}
and Hamiltonian
\begin{equation}\label{t8}
\mathcal{H}(z,p)=p\cdot z'-\mathcal{L}(z,z'(z,p))= \frac{1}{2\pi \alpha'} \, \frac{-V(z)}{\sqrt{M(z)z'^2+V(z)}}= \frac{1}{( 2\pi \alpha')^2 } \,\frac{-V(z)}{\mathcal{L}}\,.
\end{equation}
This quantity is a constant of motion, for evolution in $x$, that can be conveniently evaluated at the maximum  value of coordinate $z$: $z_{*}=z(0)$ where  $z'(0)=0$ leading to 
\begin{equation}\label{t9}
\mathcal{H}(z_{*},0)=- \frac{1}{2\pi \alpha'} \,\sqrt{V(z_{*})}\,.
\end{equation}
So, one  can express the Lagrangian as
\begin{equation}\label{t10}
\mathcal{L}=\frac{V(z)}{2\pi \alpha'\,\sqrt{V(z_{*})}}\,,
\end{equation}
and get the differential equation for the string profile:
\begin{equation}\label{t11}
\frac{dz}{dx}=\pm\frac{\sqrt{V(z)}}{\sqrt{M(z)}}\frac{\sqrt{V(z)-V(z_{*})}}{\sqrt{V(z_{*})}}\,.
\end{equation}
The distance between the infinitely massive 'quarks' is then:
\begin{equation}\label{t12}
L=\int dx=\int  \left( \frac{dz}{dx}\right)^{-1}dz=2\int_{0}^{z_{*}}\frac{\sqrt{M(z)}}{\sqrt{V(z)}}\frac{\sqrt{M(z_{*})}}{\sqrt{V(z)-V(z_{*})}}dz \,. 
\end{equation}

The on shell action of the  static string  takes the form: 
\begin{equation}\label{t12.1}
 S^{on \, shell}_{NG}= \frac{\mathcal{T}}{\pi \alpha'} \, \int^{z_{*}}_{0}\frac{\sqrt{M(z)}}{\sqrt{V(z)}}\frac{V(z)}{\sqrt{V(z)-V(z_{*})}}dz \,.
\end{equation}
The real part of the potential is obtained as the limit: 
\begin{equation}\label{limit}
Re\ V_{Q\bar{Q}} \,=\, \lim_{\mathcal{T} \to \infty } \frac{ S^{on \, shell}_{NG}}{{\mathcal{T}}} = \frac{\mathcal{T}}{\pi \alpha'} \, \int^{z_{*}}_{0}\frac{\sqrt{M(z)}}{\sqrt{V(z)}}\frac{V(z)}{\sqrt{V(z)-V(z_{*})}}dz \,.
\end{equation}

This expression is singular and is regularized by the subtraction of the quark masses: 
\begin{equation}\label{t13}
m_{Q}= \frac{1}{2\pi \alpha'} \,\int_{0}^{\infty}  \sqrt{M(z)}dz \,.
\end{equation}
The regularized form of the real part of the potential is:
\begin{equation}\label{t14}
Re\ V_{Q\bar{Q}}^{reg} \,
= \, \frac{1}{\pi \alpha'}  \int_{0}^{z_{*}}\frac{\sqrt{M(z)}}{\sqrt{V(z)}}\frac{V(z)}{\sqrt{V(z)-V(z_{*})}}dz- \frac{1}{\pi \alpha'} \,\int_{0}^{\infty} \sqrt{M(z)}dz\,.
\end{equation}
Now we discuss the imaginary part of the potential.  We follow at this point ref.  \cite{Finazzo:2013aoa} where one calculates the fluctuations of the string that cross the horizon leading to imaginary contributions  to the energy. 
The calculations of this reference were performed using the radial coordinate  $U=R^2/z$. Here, with the purpose of simplifying the  description of  the string profile in the soft wall background, we use coordinate $z$. 
 In order to  consider the same kind of fluctuations of
the metric world sheet and find a result that can be compared to this reference, we consider 
for the fluctuations of the string profile the coordinate $U$. 
 
To extract the imaginary part of the quark anti-quark potential one considers  the effect of  thermal worldsheet fluctuations  about the classical configurations $U=U_{c}(x) = R^2/z_{c}$. Fluctuation of the form 
\begin{equation}\label{t20}
U(x)=U_{c}(x)\rightarrow U(x)=U_{c}(x)+\delta U(x)\,,
\end{equation}
produce negative contributions to the root square that appears in the Nambu Goto string action of  eq.  (\ref{t3}) near $x=0$ and  generate an imaginary part in effective string action. 
Considering the long wavelength limit the fluctuations $\delta U(x) $ at each string point are independent functions.
The condition of fixed endpoits is: $\delta U(\pm L/2)=0$.

The string partition function that takes into account  the fluctuations is then a functional integral over the contributions coming from $  S_{NG}(U_{c}(x)+\delta U(x))$.  One  discretizes the interval $-L/2<x < L/2$ by considering $2N$ points located at coordinates $x_{j}=j\Delta x$ $(j=-N,-N+1,...,N)$ with $\Delta x \equiv L/(2N)$. The continuum limit  $N\rightarrow \infty$ is taken at the end of calculation. Then, $Z_{str}$ becomes
\begin{equation}\label{Action}
Z_{str}\sim \lim_{N \to \infty } \int d[\delta U(x_{-N})]...d[\delta U(x_{N})]exp \left[\frac{\mathcal{T}\Delta x}{2\pi \alpha'}\sum_{j}\sqrt{M(U_{j})(U'_{j})^2+V(U_{j})}\right],
\end{equation}
where $U_{j}\equiv U(x_{j})$ and $U'_{j}\equiv U'(x_{j})$. The thermal fluctuations are more important around $x=0$, where $U=U_{*}$ and the string is closer to the horizon.  Thus, it is reasonable to expand $U_{c}(x_{j})$ around $x=0$ and keep only terms up to second order in $x_{j}$. Given that  $U'_{c}(0) =0$  one has:
\begin{equation}
\label{expansion}
U_{c}(x_{j})\approx U_{*}+\frac{x_{j}^2}{2}U''_{c}(0).
\end{equation}
The corresponding expansion for the relevant quantities  $V(U)$ and $M(U)$,  keeping only the term up to second order in the monomial $x^{m}_{j} \delta U_{n}$ (that means $m+n \le 2$) reads
\begin{eqnarray}
\label{t24}
V(U_{j}) &\approx&  V_{*}+\delta U V'_{*}+U''_{c}(0)V'_{*}\frac{x^{2}_{j}}{2}+\frac{\delta U ^{2}}{2}V''_{*}\cr
M(U) &\approx &M(U_{*}) \,,
\end{eqnarray}
where $V_{*}\equiv V(U_{*})$,$V'_{*}\equiv V'(U_{*})$, etc.     So, one can approximate the exponent in Eq.(\ref{Action}) as
\begin{equation}\label{stringaction}
S^{NG}_{j}=\frac{\mathcal{T}\delta x}{2 \pi \alpha'}\sqrt{C_{1}x_{j}^{2}+C_{2}},
\end{equation}
with
 
\begin{equation}\label{t26}
C_{1}=\frac{U''_{c}(0)}{2}\left[2 M_{*}U''_{c}(0)+V'_{*} \right]\,\,;\,\,C_{2}=V_{*}+\delta U V'_{*}+\frac{\delta U^2}{2}V''_{*}.
\end{equation}

If the function in the square root of eq.(\ref{stringaction}) is negative then $S_{j}^{j}$ contributes to an imaginary part in the potential. The relevant region of the fluctuations is the one between the values of 
$\delta U $ that lead to a vanishing argument in the square root in the action (\ref{stringaction}). 
So, one can isolate the j-th contribution
\begin{equation}\label{relevant}
I_{j}\equiv \int^{\delta U_{jmax}}_{\delta U_{jmin}}d(\delta U_{j})exp\left[i\frac{\mathcal{T}\Delta x}{2 \pi \alpha'}\sqrt{C_{1}x_{j}^{2}+C_{2}}\right],
\end{equation}  
where $\delta U_{jmin}$, $\delta U_{jmax}  $ are the roots of $ C_{1}x_{j}^{2}+C_{2} $ in  $\delta U$.  
  
  
The integral in eq.(\ref{relevant}) can be evaluated  using the saddle point method in the classical gravity approximation  where $\alpha'\ll1$. The exponent has a stationary point when the function inside the root square of eq. (\ref{relevant})
\begin{equation}\label{t29}
D(\delta U_{j})\equiv C_{1}x_{j}^{2}+C_{2}(\delta U_{j}),
\end{equation} 
assumes an extremal value. This happens for
\begin{equation}\label{t30}
\delta U=-\frac{V'_{*}}{V''_{*}}.
\end{equation} 

Requiring  that the square root has an imaginary part implies that
$$ D(\delta U_{j})<0\rightarrow-x_{c} < x_{j} <  x_{c} \,,$$ 
where
  
\begin{equation}\label{t31}
x_{c}=\sqrt{\frac{1}{C_{1}}\left[\frac{V'^{2}_{*}}{2 V''_{*}} \right]}.
\end{equation}
We take $x_{c}=0$ if the square root in Eq.(\ref{t31}) is not real. Under these conditions, we can approximate $D(\delta U)$ by $D(\frac{-V'_{*}}{V''_{*}})$  in   eq.(\ref{relevant})

\begin{equation}\label{t32}
I_{j}\sim exp\left[i\frac{\mathcal{T}\Delta x}{2 \pi \alpha'}\sqrt{C_{1}x_{j}^{2}+V_{*}-\frac{V'_{*}}{2V''_{*}}}\right].
\end{equation}

The total contribution  to the imaginary part  comes from superposing the individual terms:  $\Pi_{j}I_{j}$. The result is\cite{Finazzo:2013aoa}
 \begin{equation}\label{t33}
Im\ V_{Q\bar{Q}}=-\frac{1}{2 \pi \alpha'}\int_{|x|<x_{c}}dx\sqrt{-x^2 \, C_{1}-V_{*}+\frac{V'^{2}_{*}}{2V''_{*}}}.
\end{equation}
After integrating over the string spatial parameter $x$ one finds: 
\begin{equation}\label{t34}
Im\ V_{Q\bar{Q}}=-\frac{1}{2\sqrt{2}\alpha'}\sqrt{M_{*}}\left[\frac{V'_{*}}{2V''_{*}}-\frac{V_{*}}{V'_{*}}\right]\,.
\end{equation}

Changing variables back to the coordinate $z = R^2/U$, that is more appropriate for working in the soft wall background,
the result for the imaginary part of the potential reads:
\begin{equation}\label{t17}
Im \ V_{Q\bar{Q}}=-\frac{1}{2\sqrt{2}\alpha'}\sqrt{M(z_{*})}\frac{z_{*}^2}{R^{2}}\left[\frac{V(z_{*})}{z_{*}^{2}V'(z_{*})}-\frac{z_{*}^{2}V'(z_{*})}{4z_{*}^{3}V'(z_{*})+2z^{4}V''(z_{*})}\right].
\end{equation}

This expression for the imaginary part of the potential is valid for metrics of the form given in eq. (\ref{t4}) with $M$ and $V$ defined in eqs. (\ref{M}) and (\ref{V}) respectively. The root of $G_{00} (z) = 0$ defines the horizon position $z_h$. We assume
that  $ \lim_{z\to z_h} \left(  G_{00}G_{zz} \right) $ is finite.  It is important to remark that the approximations used depend on the second derivative  $V''_{*}$ with respect to of coordinate $U$ in eq. (\ref{t31}) been different from zero\cite{Noronha:2009da}.


The metric of the soft wall model at finite temperature is
\begin{equation}\label{t18}
ds^2=\frac{R^2}{z^2}e^{\frac{c}{2}z^{2}}(f(z)dt^2+d\vec{x}^2+\frac{dz^2}{f(z)})\,,
\end{equation}
where $f(z)=1-z^4/z_{h}^{4}$ and  the horizon position is related to the gauge theory temperature by:   $T=1/(\pi z_{h})$.
The dilaton exponential factor $ e^{\frac{c}{2}z^{2}}$ is chosen with the positive sign that provides confinement at low temperatures \cite{Andreev:2006nw}.  The parameter $\sqrt{c} $ represents an infrared energy scale. 
 In the case of $T=0$  the masses grow linearly with the radial excitation number  \cite{Karch:2006pv,Andreev:2006vy}. 
This fact allows us to fix the value of c from the $\rho$ meson trajectory. From \cite{Andreev:2006vy} one has
  $ c \approx 0.9 \ GeV^2 $. We will use this value here. 

For this metric one finds:
\begin{eqnarray}
M(z)&=&\frac{R^4}{z^4}e^{cz^{2}} \cr
V(z)&=&R^4e^{cz^{2}}\left(\frac{1}{z^4}-\frac{1}{z_{h}^4} \right)\,,
\label{MVsoftwall}
\end{eqnarray}
so that the classical solution, given in eqs. (\ref{t12}) and  (\ref{t12.1})  leads to
\begin{equation}\label{Lsoftwall}
LT=\frac{2}{\pi}h\sqrt{1-h^4}\int_{0}^{1}dy \frac{y^2}{\sqrt{(1-y^4h^4)}\sqrt{e^{cz_{*}^{2}(y^{2}-1)}(1-y^{4}h^{4})-(1-h^4)y^4)}}\,,
\end{equation}
and
\begin{equation}\label{REsoftwall}
\frac{Re \ V_{Q\bar{Q}}}{T}= -\frac{1}{h}+\frac{1}{h}\int_{0}^{1} dy \frac{1}{y^2} \left[ \frac{e^{\frac{c}{2}z_{*}^2y^2}\sqrt{1-h^4y^4}}{\sqrt{(1-h^4y^4)-e^{\frac{c}{2}z_{*}^2y^2}y^4(1-h^4)}} -1 \right]\,,
\end{equation}
where $h=z_{*}/z_{h}$.

The imaginary part of the potential, from eq. (\ref{t17}),  takes the form:
\begin{equation}\label{ImSoftwall}
\frac{Im \ V_{Q\bar{Q}}}{T}=-\frac{3\pi}{2\sqrt{2}}\frac{e^{\frac{c}{2}z_{*}^2}}{h}\left[\frac{(2-cz_{*}^2(1-h^4)}{(-16cz_{*}^{2}+(4(cz_{*}^{2})^2+6cz_{*}^2)(1-h^4)+12))}-\frac{(1-h^4)}{2(2-cz_{*}^{2}(1-h^4))}\right].
\end{equation}

In the case of the soft wall model there are two different thermal phases. As discussed in \cite{Andreev:2006nw},  there is a critical
temperature related to the soft wall infrared scale:
$$ T_c^2 = \frac{ c}{   \pi^2 \sqrt{27} }\,  \approx \, ( 140 MeV )^ 2. $$ 
For temperatures below $T_c$ the model is confining, meaning that a quark anti-quark potential has a linear term. For temperatures above $T_c$ the model is in a nonconfined phase, representing a plasma. We are interested in describing the quarkonium state inside the plasma, so that we will consider only temperatutes above $T_c$.  Furthermore, for temperatures below $T_c$ the imaginary part of the potential 
would be zero, as discussed in \cite{Finazzo:2013aoa}. 

We introduce the dimensionless parameter $\gamma=T/T_{c}$ to characterize the temperature of the medium
and consider the region   $\gamma>1$.  Then, we can rewrite the expressions  (\ref{Lsoftwall}),(\ref{REsoftwall}) and (\ref{ImSoftwall}) as
\begin{equation}\label{Lsoft}
LT=\frac{2}{\pi}h\sqrt{1-h^4}\int_{0}^{1}dy \frac{y^2}{\sqrt{(1-y^4h^4)}\sqrt{e^{\frac{\sqrt{27}}{\gamma}^2 h^2(y^{2}-1)}(1-y^{4}h^{4})-(1-h^4)y^4)}},
\end{equation}

\begin{equation}\label{Realsoft}
\frac{Re \ V_{Q\bar{Q}}}{T}= -\frac{1}{h}+\frac{1}{h}\int_{0}^{1} dy \frac{1}{y^2} \left[ \frac{e^{\frac{\sqrt{27}}{2\gamma^2}h^2y^2}\sqrt{1-h^4y^4}}{\sqrt{(1-h^4y^4)-e^{\frac{\sqrt{27}}{\gamma^2}h^2y^2}y^4(1-h^4)}} -1 \right]
\end{equation}
and

\begin{eqnarray}
\label{Imagsoft}
\frac{Im \ V_{Q\bar{Q}}}{T} &=& -\frac{3\pi}{2\sqrt{2}}\frac{e^{\frac{1}{2}\frac{\sqrt{27}}{\gamma^2}h^2}}{h}\,\large\lbrack \frac{(2-\frac{\sqrt{27}}{\gamma^2}h^2(1-h^4)}{(-16cz_{*}^{2}+(4(\frac{\sqrt{27}}{\gamma^2}h^2)^2+6\frac{\sqrt{27}}{\gamma^2}h^2)(1-h^4)+12))} \cr
& -& \frac{(1-h^4)}{2(2-\frac{\sqrt{27}}{\gamma^2}h^2(1-h^4))}\Large]
\end{eqnarray}

\section{Thermal Width in the soft wall model }

\subsection{Review of the black hole case} 

For the sake of comparison, let us first briefly review how the thermal width was calculated in ref. \cite{Finazzo:2013aoa}. In this reference the background describing the quark gluon plasma is just an AdS black hole in Poincare coordinates. This geometry is dual to a  gauge theory, with no mass scale.   The strategy in this reference was to get the thermal width 
 from the expectation value of the imaginary part of the potential in a state of a quark anti-quark pair in 
  non-relativistic approximation
\begin{equation}\label{t35a}
\Gamma_{Q\bar{Q}}=- \langle \psi \vert  Im V_{Q\bar{Q}}  (L,T) \vert \psi \rangle \,,
\end{equation}
where 
\begin{equation}\label{t36}
\langle \vec{r} \vert \psi \rangle = \frac{1}{\sqrt{\pi}a_{0}^{3/2}}e^{-r/a_{0}}\,,
\end{equation}
is the ground-state wave function  of a particle in a Coulomb-like potential of the form $V(L)=-D/L$ and $a_{0}=2/(m_{Q}D)$ is  the Bohr radius.  In this way, a mass scale was introduced in the problem through the parameter $ a_0$ that is related to heavy quark masss $m_{Q} = 4.7 GeV $. It is interesting to mention that an expression similar to eq.(\ref{t36}) was obtained 
in ref.{\cite{Papadimitriou:2013jca}  from a two point correlator obtained holographically.

\begin{figure}[h]
\label{g6}
\begin{center}
\includegraphics[scale=0.6]{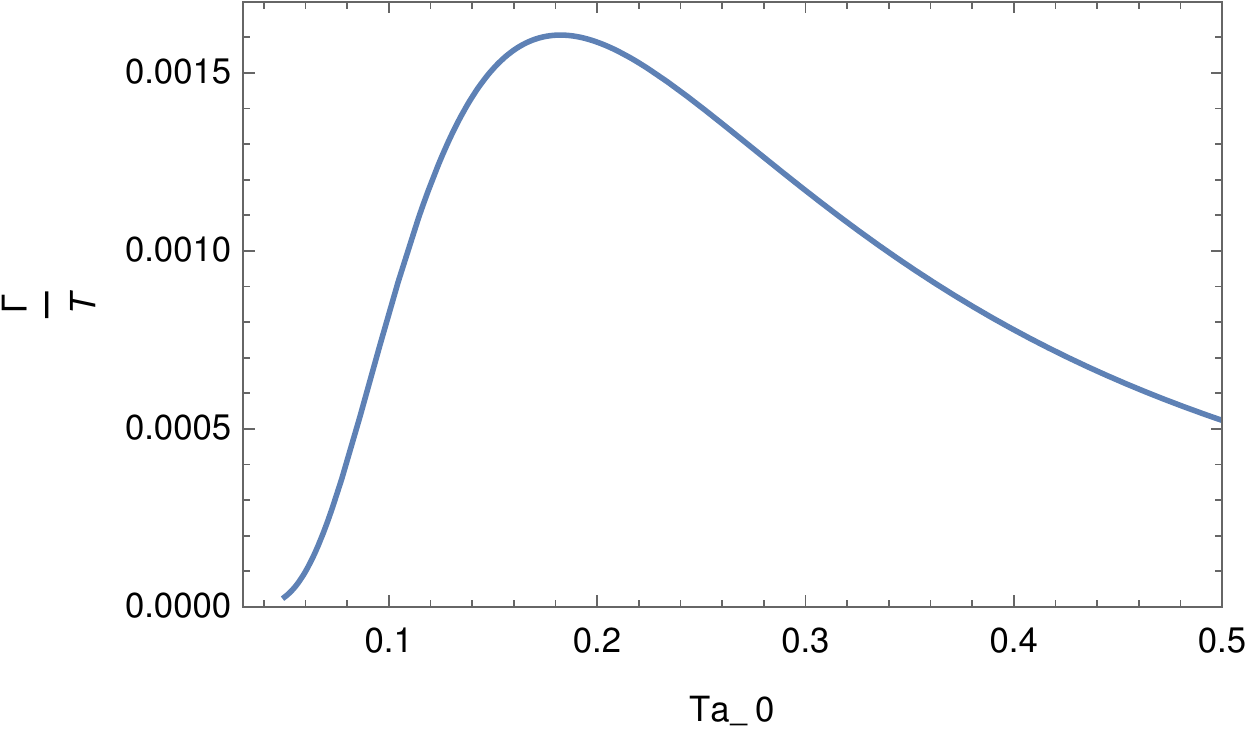}
\end{center}
\caption{Thermal width obtained in \cite{Finazzo:2013aoa}  using an AdS black hole metric. }
\end{figure}

The thermal width is then given by 
\begin{equation}\label{t39}
\frac{\Gamma}{T }\, =\, -\frac{4}{(a_{0} T)^{3}}\int   d(LT) \ ( L T)^2 \ e^{\frac{-2LT}{a_{0}T }} \, \frac{ Im\ V_{Q\bar{Q}}(L,T)}{T}
\end{equation}


This integral is performed in the interval of $LT$ where the string solutions with ``U-shaped''  profile exist.
As discussed in refs. \cite{Rey:1998bq,Brandhuber:1998bs} these solutions exist in the black hole AdS space 
for $LT \le  (LT)_{max} \approx 0.276 $.
On the other hand,  one has to impose also a lower limit for  $LT$:  $(LT)_{min}  \approx 0.266 $ since for smaller values of $LT$ the expression (\ref{t31}) for $x_c$ becomes imaginary. The quantity $x_c$ defines the interval where the coordinate $xj$ is defined, so it must be real otherwise the approximation used would not be valid. 
So, we integrate in the interval  $(LT)_{min}  \le LT  \le (LT)_{max}  $. This leads to a negative imaginary part for the potential.  
The result for the thermal width as a function of $a_0T $ is shown in figure {\bf 1} 

\subsection{ Soft wall results} 

Now, returning to our soft wall case,  the  background already contains a mass scale, the soft wall parameter $\sqrt{c} $ associated with the string tension.  So, we follow a different strategy. We calculate the expectation value of the imaginary potential by integrating over the string lengths using the semiclassical approximation: 

\begin{equation}\label{t30}
	\Gamma_{Q\bar{Q}} \,=\,- \langle Im  V_{Q\bar{Q}} (\hat{L}) \rangle_{T}=  - \frac{\int dL \ e^{-S_{NG}(L,T)} Im \ V_{Q\bar{Q}}(L,T)}{ \int dL \ e^{-S_{NG}(L,T)} } \,
\end{equation}
where $S_{NG}$ is the Nambu-Goto action with the soft wall background.  

In the desconfined regime we can approximate  the Nambu Goto action in a Coulomb form for  $LT <<  1$. Then,  using the fact that the time integral in this Euclidean metric gives a factor of the inverse of the temperature,  we can rewrite the equation in the dimensionless form as a function of $\gamma = T/T_c$
\begin{equation}
\label{Gamma}
\frac{\Gamma_{Q\bar{Q}}}{T}(\gamma)= -\frac{\int dw \ e^{D/w(\gamma)} \  \frac{Im \ V_{Q\bar{Q}}}{T}(\gamma)}{\int dw \ e^{D/w(\gamma)} } \,
\end{equation}
where $w=LT$ and $ D = 4\pi^2 \sqrt{\lambda} / \Gamma(1/4)^4 \approx 0.66 $.

\begin{figure}[h]
\label{g6}
\begin{center}
\includegraphics[scale=0.6]{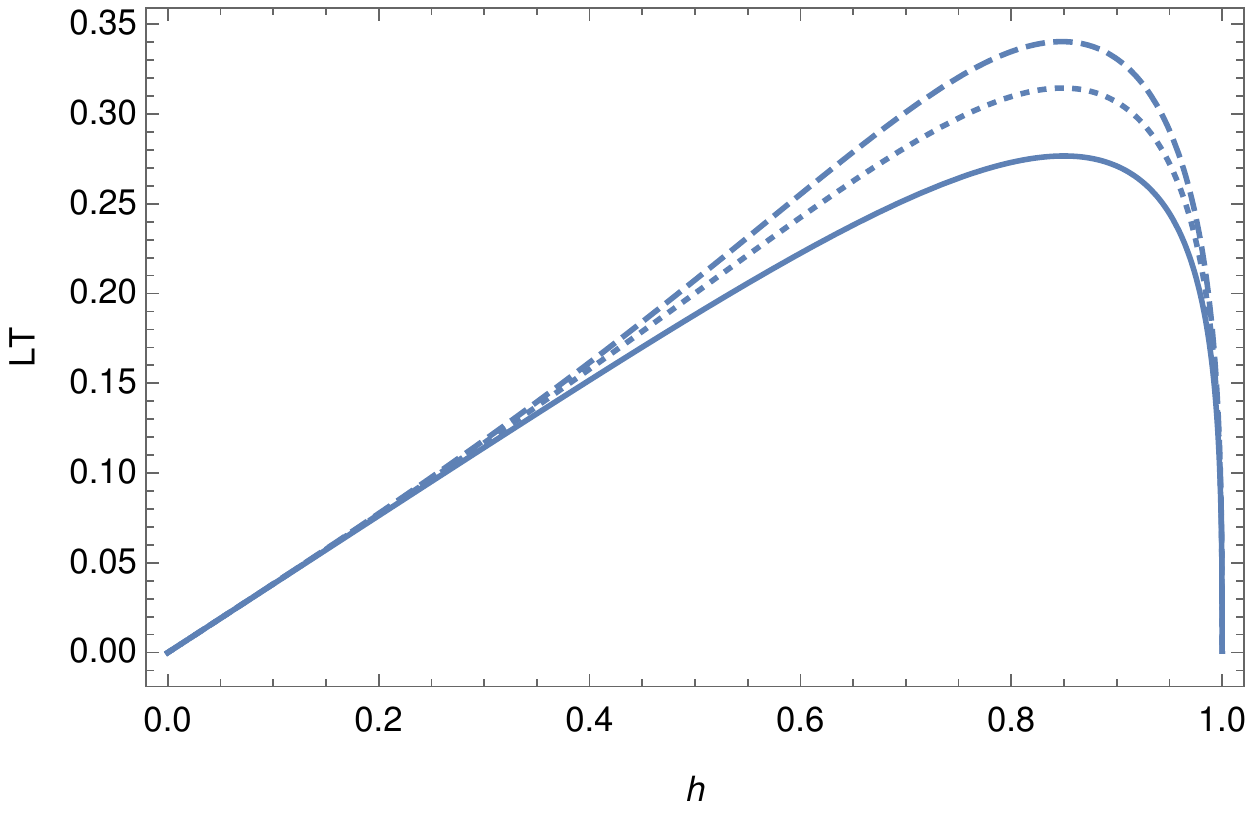}
\end{center}
\caption{The quark distance LT   versus the parameter $h = z_*/z_h$  for  $\gamma=\infty$ (solid line), $\gamma=2.5$ (dotted line)  and $\gamma=2$ (dashes line) for the soft wall model. }
\end{figure}


 Now let us discuss what are the limits of integration in the  string length $L$ that should be used in eq.(\ref{Gamma}).  
 As it happens in the case of eq.(\ref{t39}) corresponding to the black hole AdS space without soft wall,  in order to have a consistent procedure, the quantity $x_c$ must be real. This restricts our integration in the numerator of eq. (\ref{Gamma})  to a lower limit $ w = LT > (LT)_{min} $. But now, in the presence of the soft wall, the minimum value depends on the parameter $\gamma = T/T_c$ as can be seen in eq. (\ref{Lsoft}).
 The integrals  in eq. (\ref{Gamma})  must also have an upper limit, since the string ``U-shaped'' profile considered has a maximum value of L, as discussed in refs. \cite{Rey:1998bq,Brandhuber:1998bs}.  
 We show in figure {\bf 2}  the maximum values of  $LT$  as a function of $h = z_*/z_h $, obtained numerically, for three different values of $ \gamma$. 
  
In the case of the denominator we restrict the lower limit of integration considering that the quarks have a finite (large) mass, so they should  lie on a D7-Brane at the position  $z = z_{D7}$. The mass of the quark is related to the position of the D7 brane \cite{Kruczenski:2003be}:
\begin{equation}\label{t34}
m_{q}=\frac{R^2}{2 \pi \alpha' z_{D7}} \,.
\end{equation}
We can fix the mass the brane position using  the mass of the bottom quark   $m_{b}=4.7 \ GeV$  and choosing $ R^2/\alpha'=\sqrt{\lambda}=3$   to find  
 \begin{equation}\label{t35}
\frac{1}{ z_{D7}} \,=\, 9.82 \ GeV
\end{equation}
We use this value to find numerically that the lower limit of the normalization integral of the denominator is $ LT = 0.04 $.

Finally we estimate the thermal width using the equations (\ref{Lsoft}) and (\ref{Imagsoft}) for different values of $\gamma$ to calculate the thermal width in eq. (\ref{Gamma}). 
We present in figure {\bf 3} our result for the thermal width as a function of $T/T_c$. 
 Note that the thermal width is zero when $\gamma=T/T_{c}=1$  because our model is confining for lower temperatures. 
 For higher temperatures, there is a plasma. The thermal width increases in the region up to  $ T \approx 1.2 T_{c} $. For temperatures higher than $T \approx 1.2 T_{c}$   the thermal width decreases. This behaviour is qualitatively similar to   that found in \cite{Finazzo:2013aoa}.  
 
 \section{Conclusions}
  
The gauge/gravity duality  is an interesting tool to study the imaginary part of the heavy quark potential in strongly coupled plasma. This imaginary part can be used to calculate the thermal width of heavy quarkonia in such a thermal medium. 
In ref. \cite{Noronha:2009da} a method for describing  thermal worldsheet fluctuations was developed.
Then this approach was used in \cite{Finazzo:2013aoa} to obtain a lower bound for the thermal width of heavy quarkonium states in AdS black hole and also Gauss Bonnet gravity. In this previous study, the plasma is assumed to be isotropic and conformal.  The same method was applied  in \cite{Giataganas:2013lga,Fadafan:2013bva}  to extract the imaginary part of the heavy quark potential  but in a  strongly coupled anisotropic plasma.  

In the present work,  we considered a non-conformal strongly coupled plasma. We used  the AdS/QCD soft wall model  that 
carries  an infrared mass scale and introduces confinement in gauge gravity duality. 
This background represents a gauge theory that is a confining at low temperatures as has a deconfining transition at a critical temperature.  Another point that is different from the approach of \cite{Finazzo:2013aoa} is that in this reference the width is calculated using a wave function of a quarkonium state, while here we obtain the width by averaging over the string lengths. 
 Consistently, the result of the approach developed here is qualitatively similar to the one of reference\cite{Finazzo:2013aoa}. 
 Our result is also qualitatively similar to what is found using lattice QCD in ref. \cite{Aarts:2011sm}.  
As a final remark, it is interesting to  mention  that our procedure was developed using a different radial coordinate ($z$ instead of $U$) that is more convenient when working with the soft wall model.

 \begin{figure}\label{g6}
\begin{center}
\includegraphics[scale=0.6]{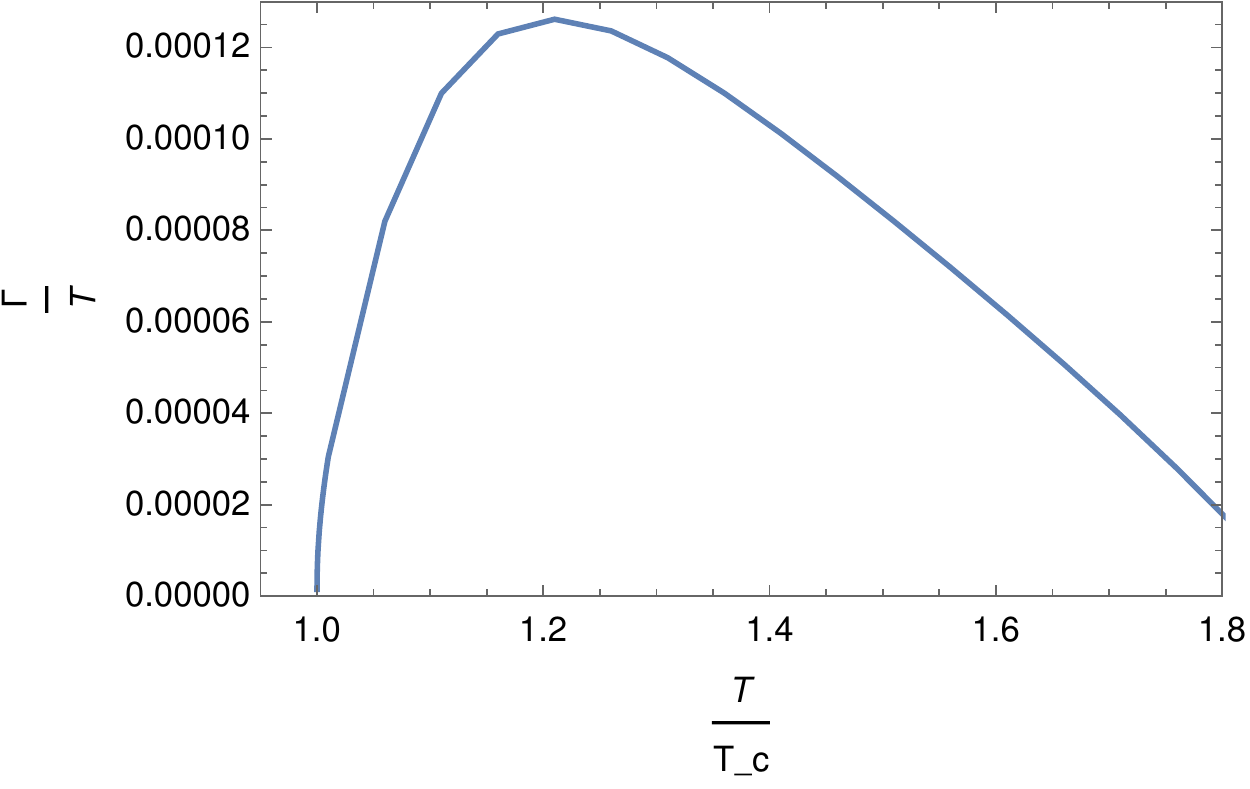}
\end{center}
\caption{Thermal width in the soft wall model }
\end{figure}

\noindent {\bf Acknowledgments:}   N.B. is partially supported by CNPq and L. F. Ferreira is supported CAPES.

 \end{document}